\documentclass[journal]{IEEEtran}
\usepackage{amsmath}
\usepackage{subfigure}
\usepackage{array}
\usepackage{algpseudocode,algorithm}
\usepackage{color}
\usepackage{cases}
\usepackage{amsfonts}
\usepackage{graphicx,color}

\renewcommand{\Vec}[1]{\textrm{\boldmath $#1$}} 
\def\pt#1{\left(#1\right)} 
\def\br#1{\left[#1\right]} 

\begin{document}
\title{
  Statistical Parametric Speech Synthesis Incorporating Generative Adversarial Networks
}

\author{Yuki~Saito,~\IEEEmembership{NonMember,}
        Shinnosuke~Takamichi,~\IEEEmembership{Member,~IEEE}
        and~Hiroshi~Saruwatari,~\IEEEmembership{Member,~IEEE}
\thanks{
The authors are with
Graduate School of Information Science and Technology,
The University of Tokyo,
7-3-1 Hongo, Bunkyo-ku, Tokyo 113-8656, Japan
(e-mail: \{ yuuki\_saito, shinnosuke\_takamichi, hiroshi\_saruwatari \}@ipc.i.u-tokyo.ac.jp).
}
\thanks{
Part of this work was supported by
ImPACT Program of Council for Science,
Technology and Innovation (Cabinet Office, Government of Japan),
SECOM Science and Technology Foundation,
and JSPS KAKENHI Grant Number 16H06681
and JP17H06101.
}
}

\markboth{PREPRINT MANUSCRIPT OF IEEE/ACM TRANSACTIONS ON AUDIO, SPEECH AND LANGUAGE PROCESSING \copyright 2017 IEEE \hfill}%
{Shell \MakeLowercase{\textit{et al.}}: Bare Demo of IEEEtran.cls for IEEE Journals}

\maketitle

\begin{abstract}
  A method for statistical parametric speech synthesis
  incorporating generative adversarial networks (GANs) is proposed.
  Although powerful deep neural networks (DNNs) techniques can be applied to
  artificially synthesize speech waveform,
  the synthetic speech quality is low
  compared with that of natural speech.
  One of the issues causing the quality degradation is
  an over-smoothing effect often observed in the generated speech parameters.
  A GAN introduced in this paper
  consists of two neural networks:
  a discriminator to distinguish natural and generated samples,
  and a generator to deceive the discriminator.
  In the proposed framework incorporating the GANs,
  the discriminator is trained to distinguish
  natural and generated speech parameters,
  while the acoustic models are trained to minimize
  the weighted sum of the conventional minimum generation loss and
  an adversarial loss for deceiving the discriminator.
  Since the objective of the GANs is to minimize
  the divergence (i.e., distribution difference)
  between the natural and generated speech parameters,
  the proposed method effectively alleviates the over-smoothing effect on
  the generated speech parameters.
  We evaluated the effectiveness for
  text-to-speech and voice conversion,
  and found that
  the proposed method can generate more natural spectral parameters and $F_0$
  than conventional minimum generation error training algorithm
  regardless its hyper-parameter settings.
  Furthermore,
  we investigated the effect of the divergence of various GANs,
  and found that
  a Wasserstein GAN minimizing the Earth-Mover's distance works
  the best in terms of improving synthetic speech quality.
\end{abstract}


\IEEEpeerreviewmaketitle

\section{Introduction}
\IEEEPARstart{S}{tatistical} parametric speech synthesis (SPSS)~\cite{zen09}
is a technique
that aims to generate natural-sounding synthetic speech.
Text-to-speech (TTS) synthesis~\cite{sagisaka88}
is a technique for synthesizing speech from text,
and voice conversion (VC)~\cite{stylianou88}
is a technique for synthesizing speech
from another one
while preserving linguistic information of original speech.
In SPSS,
acoustic models represent the relationship
between input features and acoustic features.
Recently,
deep neural networks (DNNs)~\cite{ling15} have been utilized
as the acoustic models for TTS and VC
because they can model the relationship between input features and
acoustic features more accurately than conventional
hidden Markov models~\cite{tokuda13}
and Gaussian mixture models~\cite{toda07_MLVC}.
These acoustic models are trained with several training algorithms
such as the minimum generation error (MGE) criterion~\cite{wu06mgehmm, wu16mge}.
Techniques for training the acoustic models to generate high-quality speech
are widely studied since they can be used for both TTS and VC.
However,
the speech parameters generated from these models
tend to be over-smoothed, and
the resultant quality of speech is
still low compared with that of natural speech~\cite{zen09, toda16vc}.
The over-smoothing effect is a common issue
in both TTS and VC.

One way to improve speech quality is
to reduce the difference between natural and generated speech parameters.
For instance,
since the parameter distributions of natural and synthetic speech
are significantly different~\cite{ijima16},
we can improve the synthetic speech quality by
transforming the generated speech parameters
so that their distribution is close to that of natural speech.
This can be done by, for example,
modeling the probability distributions in a parametric~\cite{toda07_MLVC}
or non-parametric~\cite{ohtani12histogram} way in the training stage,
and then,
generating or transforming the synthetic speech parameters
by using the distributions.
The more effective approach is to use analytically derived features
correlated to the quality degradation of the synthetic speech.
Global variance (GV)~\cite{toda07_MLVC} and
modulation spectrum (MS)~\cite{takamichi16mspf} are
well-known examples for reproducing natural statistics.
These features work as a constraint
in the training/synthesis stage~\cite{takamichi15_icassp_trj, hashimoto16}.
Nose and Ito~\cite{nose14_enhance_gv}
and Takamichi et al.~\cite{takamichi15_icassp_trj} proposed methods
that reduce the difference between the Gaussian distributions of
natural and generated GV and MS.
However,
quality degradation is still a critical problem.

In order to address this quality problem,
in this paper
we propose a novel method using
generative adversarial networks (GANs) for
training acoustic models in SPSS.
A GAN consists of two neural networks:
a discriminator to distinguish natural and generated samples,
and a generator to deceive the discriminator.
Based on the framework,
we define a new training criterion for the acoustic models;
the criterion is the weighted sum of the conventional MGE training
and an adversarial loss.
The adversarial loss makes the discriminator
recognize the generated speech parameters as natural.
Since the objective of the GANs is to minimize the divergence
(i.e., the distribution difference)
between the natural and generated speech parameters,
our method effectively alleviates the effect of over-smoothing
the generated speech parameters.
Moreover,
our method can be regarded as
a generalization of
the conventional method using explicit modeling of
analytically derived features such as GV and MS
because
it effectively minimizes the divergence without explicit statistical modeling.
Also,
the discriminator used in our method can be interpreted as anti-spoofing,
namely,
a technique for detecting synthetic speech and preventing voice spoofing attack.
Accordingly, techniques and ideas concerning anti-spoofing
can be applied to the training.
We evaluated the effectiveness of the proposed method in
DNN-based TTS and VC,
and found that
the proposed algorithm generates more natural spectral parameters and $F_0$
than those of the conventional MGE training algorithm
and improves the synthetic speech quality
regardless its hyper-parameter settings
which control the weight of the adversarial loss.
Furthermore,
we investigated the effect of the divergence of
various GANs,
including image-processing-related ones
such as the least squares GAN (LS-GAN) and the Wasserstein GAN (W-GAN),
and speech-processing-related ones such as the $f$-divergence GAN ($f$-GAN).
The results of the investigation demonstrate that
the W-GAN minimizing the Earth-Mover's distance works
the best in regard to improving synthetic speech quality.

In Section II of this paper,
we briefly review conventional training algorithms
in DNN-based TTS and VC.
Section III introduces GANs and proposes a method for speech synthesis incorporating those GANs.
Section IV presents the experimental evaluations.
We conclude in Section V with a summary.

\section{Conventional DNN-based SPSS}
This section describes the conventional training algorithm for
DNN-based SPSS,
including TTS and VC.

\subsection{DNN-based TTS}
\subsubsection{DNNs as Acoustic Models}
In DNN-based TTS~\cite{zen13dnn},
acoustic models representing
the relationship between linguistic features and speech parameters
consist of layered hierarchical networks.
In training the models,
we minimize the loss function
calculated using the speech parameters of natural and synthetic speech.
Let
$\Vec{x} = [\Vec{x}_1^\top, \cdots, \Vec{x}_t^\top, \cdots, \Vec{x}_T^\top]^\top$
be a linguistic feature sequence,
$\Vec{y} = [\Vec{y}_1^\top, \cdots, \Vec{y}_t^\top, \cdots, \Vec{y}_T^\top]^\top$
be a natural speech parameter sequence,
and
$\Vec{\hat y} = [\Vec{\hat y}_1^\top, \cdots, \Vec{\hat y}_t^\top, \cdots, \Vec{\hat y}_T^\top]^\top$
be a generated speech parameter sequence,
where $t$ and $T$ denote the frame index and total frame length, respectively.
$\Vec{x}_t$
and
$\Vec{y}_t = [y_t\pt{1}, \cdots, y_t\pt{D}]^\top$
are a linguistic parameter vector
and
a $D$-dimensional speech parameter vector at frame $t$,
respectively.

\subsubsection{Acoustic model training}
The DNNs that predict a natural static-dynamic speech feature sequence
$\Vec{Y} = [\Vec{Y}_1^\top, \cdots, \Vec{Y}_t^\top, \cdots, \Vec{Y}_T^\top]^\top$
from
$\Vec{x}$
are trained to minimize a defined training criterion.
$\Vec{Y}_t = [\Vec{y}_t^\top, \Delta \Vec{y}_t^\top, \Delta \Delta \Vec{y}_t^\top]^\top$
is a natural static-dynamic speech feature at frame $t$.
Given a predicted static-dynamic speech feature sequence
$\Vec{\hat Y} = [\Vec{\hat Y}_1^\top, \cdots, \Vec{\hat Y}_t^\top, \cdots, \Vec{\hat Y}_T^\top]^\top$,
the most standard criterion is the mean squared error (MSE)
$L_{\mathrm{MSE}}(\Vec{Y}, \Vec{\hat Y})$
between
$\Vec{Y}$ and $\Vec{\hat Y}$
defined as follows:
\begin{align}
L_{\mathrm{MSE}}\pt{\Vec{Y}, \Vec{\hat Y}}
& = \frac{1}{T} \pt{\Vec{\hat Y} - \Vec{Y}}^\top \pt{\Vec{\hat Y} - \Vec{Y}}\label{eq:LMSE}.
\end{align}
A set of the model parameters $\theta_{\rm G}$
(e.g., weight and bias of DNNs)
is updated by the backpropagation algorithm using the gradient
$\nabla_{\theta_{\mathrm{G}}} L_{\rm MSE} (\Vec{Y}, \Vec{\hat Y})$.

To take the static-dynamic constraint into account,
the minimum generation error (MGE) training algorithm was proposed~\cite{wu16mge}.
In MGE training,
the loss function
$L_{\mathrm{MGE}}(\Vec{y}, \Vec{\hat y})$
is defined as the mean squared error between natural and generated speech parameters
as follows:
\begin{align}
L_{\mathrm{MGE}}\pt{\Vec{y}, \Vec{\hat y}}
& = \frac{1}{T} \pt{\Vec{\hat y} - \Vec{y}}^\top \pt{\Vec{\hat y} - \Vec{y}} \nonumber \\
& = \frac{1}{T} \pt{\Vec{R} \Vec{\hat Y} - \Vec{y}}^\top \pt{\Vec{R} \Vec{\hat Y} - \Vec{y}}\label{eq:LG}.
\end{align}
$\Vec{R}$
is a $DT$-by-3$DT$ matrix given as
\begin{align}
\Vec{R} = \pt{\Vec{W}^\top \Vec{\Sigma}^{-1} \Vec{W}}^{-1} \Vec{W}^\top \Vec{\Sigma}^{-1},
\end{align}
where
$\Vec{W}$ is a 3$DT$-by-$DT$ matrix
for calculating dynamic features~\cite{tokuda13}
and
$\Vec{\Sigma} = {\mathrm{diag}}[ \Vec{\Sigma}_1, \cdots, \Vec{\Sigma}_t, \cdots, \Vec{\Sigma}_T ]$
is a 3$DT$-by-3$DT$ covariance matrix,
where $\Vec{\Sigma}_t$ is a 3$D$-by-3$D$ covariance matrix at frame $t$.
$\Vec{\Sigma}$
is separately estimated using training data.
We define the speech parameter prediction as
$\Vec{\hat y} = \Vec{R} \Vec{\hat Y} = \Vec{G}(\Vec{x}; \theta_{\rm G})$,
where
$\theta_{\rm G}$
denotes the acoustic model parameters
and
it is updated by the backpropagation algorithm using the gradient
of the generation error,
$\nabla_{\theta_{\rm G}} L_{\rm MGE} (\Vec{y}, \Vec{\hat y})$.
As described in \cite{wu16mge},
the gradient includes
$\nabla_{\hat Y} L_{\rm MGE} (\Vec{y}, \Vec{\hat y})$
given as
$\Vec{R}^{\top} (\Vec{\hat y} - \Vec{y})/T$.

Phoneme duration is predicted in the same manner
without dynamic feature calculation.
Let
$\Vec{d} = [d_{1}, \cdots, d_{p}, \cdots, d_{P}]^\top$
be a natural phoneme duration sequence,
and
$\Vec{\hat d} = [\hat d_{1}, \cdots, \hat d_{p}, \cdots, \hat d_{P}]^\top$
be a duration sequence generated using duration models described as DNNs.
$p$ is the phoneme index and $P$ is the total number of phonemes.
The model parameters are updated to minimize $L_{\rm MSE} (\Vec{d}, \Vec{\hat d})$.

\subsection{DNN-based VC}
DNN-based acoustic models for VC
convert
input speech features to
desired output speech features.
In training,
a dynamic time warping algorithm is used to
temporally align source and target speech features.
Using the aligned features,
$\Vec{x}$
and
$\Vec{y}$,
the acoustic models are trained to minimize
$L_{\rm MGE}(\Vec{y}, \Vec{\hat y})$,
the same as DNN-based TTS.

\section{DNN-BASED SPSS INCORPORATING GAN}

\subsection{Generative Adversarial Networks (GANs)~\cite{goodfellow14}}
A GAN is a framework for learning
deep generative models,
which simultaneously trains two DNNs:
a generator and discriminator $D(\Vec{y}; \theta_{\rm D})$.
$\theta_{\rm D}$ is a set of the model parameters of the discriminator.
The value obtained by taking the sigmoid function from the discriminator's output,
$1 / (1 + \exp (-D(\Vec{y})))$,
represents the posterior probability
that input $\Vec{y}$ is natural data.
The discriminator is trained to
make the posterior probability 1 for natural data and 0 for generated data,
while
the generator is trained to deceive the discriminator;
that is,
it tries to make the discriminator make the posterior probability 1 for generated data.

\begin{figure}[!t]
  \centering
  \includegraphics[width=0.98\linewidth]{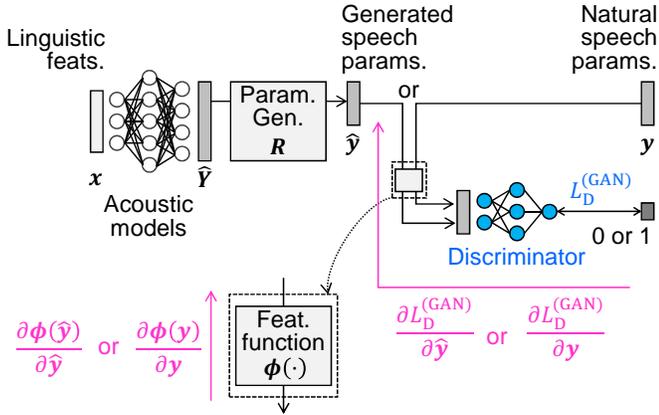}
  \vspace{-5pt}
  \caption{
    Loss function and gradients for updating the discriminator.
    Param. Gen. indicates the speech parameter generation~\cite{tokuda13}.
    Note that, the model parameters of the acoustic models
    are not updated in this step.
  }
  \label{fig:BP_D}
\end{figure}

In the GAN training,
the two DNNs are iteratively updated by minibatch stochastic gradient descent.
First,
by using
natural data $\Vec{y}$
and generated data $\Vec{\hat y}$,
we calculate the discriminator loss
$L_{\rm D}^{(\rm GAN)}\pt{\Vec{y}, \Vec{\hat y}}$
defined as the following cross-entropy function:
\begin{align}\label{eq:GAN_D}
\begin{split}
L_{\rm D}^{(\rm GAN)}\pt{\Vec{y}, \Vec{\hat y}}
&= -\frac{1}{T} \sum_{t=1}^{T} \log \frac{1}{1 + \exp \pt{ - D\pt{\Vec{y}_t} }} \\
&\quad - \frac{1}{T} \sum_{t=1}^{T} \log \pt{1 - \frac{1}{ 1 + \exp \pt{-D\pt{\Vec{\hat y}_t}}}}.
\end{split}
\end{align}
$\theta_{\rm D}$
is updated by using the stochastic gradient
$\nabla_{\theta_{\rm D}} L_{\rm D}^{(\rm GAN)}(\Vec{y}, \Vec{\hat y})$.
Figure~\ref{fig:BP_D} illustrates the procedure for computing the discriminator loss.
After updating the discriminator,
we calculate the adversarial loss of the generator
$L_{\rm ADV}^{(\rm GAN)}\pt{\Vec{\hat y}}$
which deceives the discriminator
as follows:
\begin{align}\label{eq:GAN_G}
L_{\rm ADV}^{(\rm GAN)}\pt{\Vec{\hat y}} = -\frac{1}{T} \sum_{t=1}^{T} \log \frac{1}{1 + \exp \pt{ - D\pt{\Vec{\hat y}_t} }}.
\end{align}
A set of the model parameters of the generator
$\theta_{\rm G}$
is updated by using the stochastic gradient
$\nabla_{\theta_{\rm G}} L_{\rm ADV}^{(\rm GAN)}(\Vec{\hat y})$.
Goodfellow et al.~\cite{goodfellow14}
showed this adversarial framework minimizes
the approximated Jensen--Shannon (JS) divergence between
two distributions of natural and generated data.

\subsection{Acoustic Model Training Incorporating GAN}
Here,
we describe a novel training algorithm for SPSS which incorporates the GAN.
As for the proposed algorithm,
acoustic models are trained to deceive the discriminator
that distinguishes natural and generated speech parameters.

\begin{figure}[!t]
  \centering
  \includegraphics[width=0.98\linewidth]{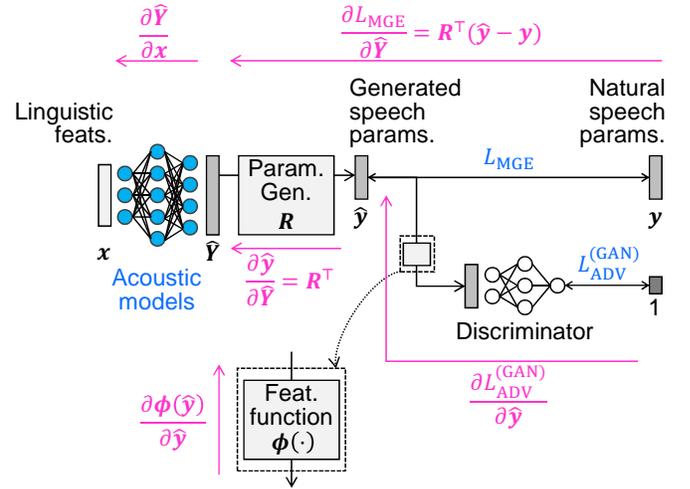}
  \vspace{-5pt}
  \caption{
    Loss functions and gradients for updating acoustic models
    in the proposed method.
    Note that the model parameters of the discriminator are not updated in this step.
  }
  \label{fig:BP_G}
\end{figure}

The loss function of speech synthesis is defined as the following:
\begin{align}\label{eq:PM}
L_{\rm G}\pt{\Vec{y}, \Vec{\hat y}}
&= L_{\mathrm{MGE}}\pt{\Vec{y}, \Vec{\hat y}} +
\omega_{\mathrm{D}}
\frac{E_{L_{\mathrm{MGE}}}}{E_{L_{\mathrm{ADV}}}}
L_{\mathrm{ADV}^{(\rm GAN)}}\pt{\Vec{\hat y}},
\end{align}
where
$L_{\mathrm{ADV}}^{(\rm GAN)}\pt{\Vec{\hat y}}$
makes the discriminator
recognize the generated speech parameters as natural,
and minimizes the divergence between the distributions of the
natural and generated speech parameters.
Therefore,
the proposed
loss function
not only minimizes the generation error
but also makes the distribution of the generated speech parameters
close to that of natural speech.
$E_{L_\mathrm{MGE}}$
and
$E_{L_\mathrm{ADV}}$
denote the expectation values of
$L_{\mathrm{MGE}}(\Vec{y}, \Vec{\hat y})$
and
$L_{\mathrm{ADV}}^{(\rm GAN)}\pt{\Vec{\hat y}}$,
respectively.
Their ratio
$E_{L_\mathrm{MGE}}/E_{L_\mathrm{ADV}}$
is the scale normalization term between the two loss functions,
and the hyper-parameter
$\omega_{\mathrm{D}}$
controls the weight of the second term.
When
$\omega_{\mathrm{D}} = 0$,
the loss function is equivalent to the conventional MGE training,
and when
$\omega_{\mathrm{D}} = 1$,
the two loss functions have equal weights.
A set of the model parameters of the acoustic models
$\theta_{\rm G}$
is updated by using the stochastic gradient
$\nabla_{\theta_{\rm G}} L_{\rm G}(\Vec{\hat y})$.
Figure~\ref{fig:BP_G} illustrates the procedure for computing the proposed loss function.
In our algorithm,
the acoustic models and discriminator
are iteratively optimized,
as shown in Algorithm~1.
\begin{algorithm}[!t]
  \caption{Iterative optimization for acoustic models and discriminator}
  \begin{algorithmic}[1]
    \State $\eta :=$ learning rate
    \For{number of training iterations}
      \ForAll{training data $(\Vec{x}, \Vec{y})$}
        \State generate $\Vec{\hat y}$ from the acoustic models:
          \[ \Vec{\hat y} = \Vec{G}(\Vec{x}).\]
        \State update $\theta_{\mathrm{D}}$ while fixing $\theta_{\mathrm{G}}$:
          \[ \theta_{\mathrm{D}} \leftarrow \theta_{\mathrm{D}} - \eta \nabla_{\theta_{\mathrm{D}}} L_{\rm D}^{(\rm GAN)}(\Vec{y}, \Vec{\hat y}).\]
        \State update $\theta_{\mathrm{G}}$ while fixing $\theta_{\mathrm{D}}$:
          \[ \theta_{\mathrm{G}} \leftarrow \theta_{\mathrm{G}} - \eta \nabla_{\theta_{\mathrm{G}}} L_{\rm G}(\Vec{y}, \Vec{\hat y}).\]
      \EndFor
    \EndFor
  \end{algorithmic}
\end{algorithm}
When one module is being updated,
the model parameters of the another are fixed;
that is,
although the discriminator is included in
the forward path to calculate
$L_{\rm ADV}^{(\rm GAN)}(\Vec{\hat y})$
in
$L_{\rm G}(\Vec{y}, \Vec{\hat y})$,
$\theta_{\rm D}$
is not updated
by the backpropagation for the acoustic models.

The discriminator used in our method can be regarded as a
DNN-based anti-spoofing (voice spoofing detection)~\cite{wu16anti_spoofing,chen15dnn_asv}
that distinguishes natural and synthetic speech.
From this perspective,
a feature function $\Vec{\phi} (\Vec{\cdot})$
can be inserted between speech parameter prediction and the discriminator
as shown in Figs.~\ref{fig:BP_D} and \ref{fig:BP_G}.
The function calculates more distinguishable features in
anti-spoofing than the direct use of speech parameters themselves.
Namely,
instead of $\Vec{y}$ and $\Vec{\hat y}$ in Eqs.~(\ref{eq:GAN_D}) and (\ref{eq:GAN_G}),
$\Vec{\phi}(\Vec{y})$ and $\Vec{\phi}(\Vec{\hat y})$ are used.
In training the acoustic models,
the gradient $\partial \Vec{\phi}(\Vec{\hat y}) / \partial \Vec{\hat y}$
is used for backpropagation.
For example,
when $\Vec{\phi}(\Vec{\hat y}) = \Vec{W}\Vec{\hat y}$,
the gradient $\Vec{W}^\top$ is used for backpropagation.

\subsection{Application to $F_0$ and Duration Generation}
Our algorithm is simply applied to the spectral parameter generation
and conversion for TTS and VC.
Here,
we extend our algorithm to $F_0$ and duration generation in TTS.
For $F_0$ generation,
we use a continuous $F_0$ sequence~\cite{yu11}
instead of the $F_0$ sequence because of the simple implementation.
The input of the discriminator is the joint vector of a spectral parameter vector
and
continuous $F_0$ value of each frame.

\begin{figure}[!t]
  \centering
  \includegraphics[width=0.9\linewidth]{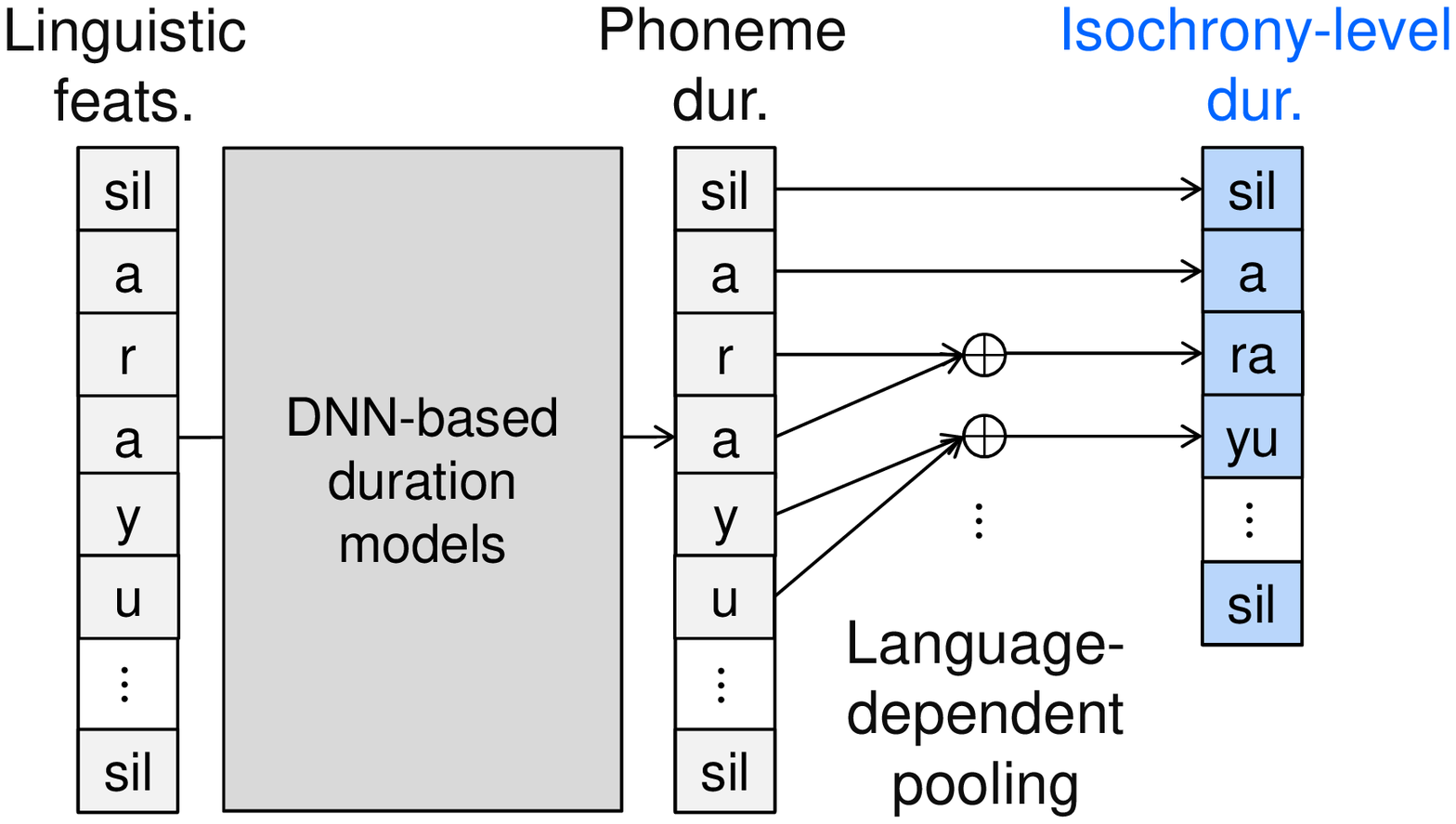}
  \vspace{-5pt}
  \caption{
    Architecture to calculate isochrony-level duration
    from phoneme duration.
    In the case of Japanese, which has mora isochrony,
    each mora duration is calculated from the corresponding phoneme duration,
    e.g.,
    the mora duration of /ra/ is calculated as
    the sum of the phoneme durations of /r/ and /a/.
  }
  \label{fig:dur}
\end{figure}

For duration generation,
although we can directly apply our algorithm to phoneme duration,
it is not guaranteed that naturally-distributed phoneme duration has
natural isochrony of the target language
(e.g., moras in Japanese)~\cite{esther02}.
Therefore,
we modify our algorithm so that the generated duration
naturally distributes in the language-dependent isochrony level.
Figure~\ref{fig:dur} shows the architecture.
In the case of Japanese,
which has mora isochrony,
each mora duration is calculated from the corresponding phoneme durations.
\begin{figure}[!t]
  \centering
  \includegraphics[width=0.925\linewidth]{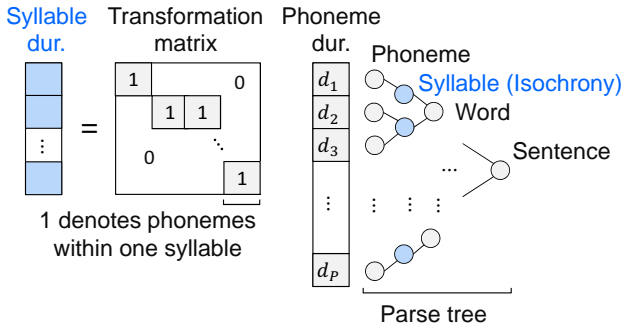}
  \vspace{-5pt}
  \caption{
    Matrix representation to calculate isochrony-level duration.
    This is an example in the case of a syllable-timed language such as Chinese.
  }
  \label{fig:calcdur}
\end{figure}
The discriminator minimizes the cross-entropy function by using the
isochrony-level duration,
while the generator minimizes the weighted sum of the
MSE between natural and generated phoneme durations
and the adversarial loss using the isochrony-level durations.
Since the calculation of the isochrony-level duration is
represented as the matrix multiplication shown in Fig.~\ref{fig:calcdur},
the backpropagation is done using the transpose of the transformation matrix.

\subsection{GANs to Be Applied to The Proposed Method}
The GAN framework works as a divergence minimization
between natural and generated speech parameters.
As described in Section III-B,
the original GAN~\cite{goodfellow14}
minimizes the approximated JS divergence.
From the perspective of the divergence minimization,
we further introduce additional GANs minimizing other divergences:
$f$-GAN~\cite{nowozin16fGAN},
Wasserstein GAN (W-GAN)~\cite{arjovsky17},
and least squares GAN (LS-GAN)~\cite{mao17}.
The divergence of the $f$-GAN is strongly related to speech processing
such as a nonnegative matrix factorization~\cite{lee00nmf,kompass07},
and the effectiveness of the W-GAN and LS-GAN in the image processing is known.
The discriminator loss $L_{\rm D}^{(*\text{-}\mathrm{GAN})}(\Vec{y}, \Vec{\hat y})$
and adversarial loss
$L_{\rm ADV}^{(*\text{-}\mathrm{GAN})}(\Vec{\hat y})$
introduced below can be used instead of
Eqs.~(\ref{eq:GAN_D}) and (\ref{eq:GAN_G}), respectively.

\subsubsection{$f$-GAN~\cite{nowozin16fGAN}}
The $f$-GAN is the unified framework that encompasses the original GAN.
The difference between distributions of natural and generated data
is defined as the $f$-divergence~\cite{csiszar04},
which is a large class of different divergences
including the Kullback--Leibler (KL) and JS divergence.
The $f$-divergence $\mathcal{D}_f(\Vec{y} \| \Vec{\hat y})$ is defined as follows:
\begin{align}\label{eq:f-div}
  \mathcal{D}_{f} \pt{\Vec{y} \| \Vec{\hat y}} = \int q \pt{\Vec{\hat y}} f \left( \frac{p\pt{\Vec{y}}}{q \pt{\Vec{\hat y}}} \right) d\Vec{y},
\end{align}
where $p(\cdot)$ and $q(\cdot)$ are
absolutely continuous density functions of $\Vec{y}$ and $\Vec{\hat y}$, respectively.
$f(\cdot)$ is a convex function satisfying $f(1) = 0$.
Although various choices of $f(\cdot)$ for recovering popular divergences are available,
we adopt ones related to speech processing.

{\bf KL-GAN:} Defining $f(r) = r \log r$ gives the KL divergence as follows:
\begin{align}\label{eq:KL-div}
  \mathcal{D}_{\rm KL} \pt{\Vec{y} \| \Vec{\hat y}} = \int p \pt{\Vec{y}} \log \frac{ p\pt{\Vec{y}} }{ q\pt{\Vec{\hat y}} } d\Vec{y}.
\end{align}
The discriminator loss
$L_{\rm D}^{(\mathrm{KL}\text{-}\mathrm{GAN})} \pt{\Vec{y}, \Vec{\hat y}}$
is defined as follows:
\begin{align}\label{eq:KLGAN_D}
\begin{split}
L_{\rm D}^{(\mathrm{KL}\text{-}\mathrm{GAN})} \pt{\Vec{y}, \Vec{\hat y}} &=
- \frac{1}{T} \sum_{t=1}^{T} D\pt{\Vec{y}_t} \\
&\quad +
\frac{1}{T} \sum_{t=1}^{T} \exp \pt{ D \pt{\Vec{\hat y}_t} -1 },
\end{split}
\end{align}
while the adversarial loss
$L_{\rm ADV}^{(\mathrm{KL}\text{-}\mathrm{GAN})} \pt{\Vec{\hat y}}$
is defined as follows:
\begin{align}\label{eq:KLGAN_G}
L_{\rm ADV}^{(\mathrm{KL}\text{-}\mathrm{GAN})} \pt{\Vec{\hat y}} &=
- \frac{1}{T} \sum_{t=1}^{T} D\pt{\Vec{\hat y}_t}.
\end{align}

{\bf Reversed KL (RKL)-GAN:}
Since the KL divergence is not symmetric,
the reversed version, called reversed KL (RKL) divergence
$\mathcal{D}_{\rm RKL}(\Vec{y} \| \Vec{\hat y})$
differs from
$\mathcal{D}_{\rm KL}(\Vec{y} \| \Vec{\hat y})$,
which is defined as follows:
\begin{align}\label{eq:RKL-div}
  \mathcal{D}_{\rm RKL} \pt{\Vec{y} \| \Vec{\hat y}} = \int q \pt{\Vec{\hat y}} \log \frac{ q\pt{\Vec{\hat y}} }{ p\pt{\Vec{y}} } d\Vec{y}
  = \mathcal{D}_{\rm KL} \pt{\Vec{\hat y} \| \Vec{y}}.
\end{align}
Defining $f(r) = - \log{r}$ gives
the discriminator loss
$L_{\rm D}^{(\mathrm{RKL}\text{-}\mathrm{GAN})} \pt{\Vec{y}, \Vec{\hat y}}$
as follows:
\begin{align}\label{eq:RKLGAN_D}
\begin{split}
L_{\rm D}^{(\mathrm{RKL}\text{-}\mathrm{GAN})} \pt{\Vec{y}, \Vec{\hat y}} &=
 \frac{1}{T} \sum_{t=1}^{T} \exp \pt{ - D\pt{\Vec{y}_t} } \\
&\quad +
\frac{1}{T} \sum_{t=1}^{T} \pt{ -1 + D \pt{\Vec{\hat y}_t}},
\end{split}
\end{align}
while the adversarial loss
$L_{\rm ADV}^{(\mathrm{RKL}\text{-}\mathrm{GAN})} \pt{\Vec{\hat y}}$
is defined as follows:
\begin{align}\label{eq:RKLGAN_G}
L_{\rm ADV}^{(\mathrm{RKL}\text{-}\mathrm{GAN})} \pt{\Vec{\hat y}} &=
 \frac{1}{T} \sum_{t=1}^{T} \exp \pt{- D\pt{\Vec{\hat y}_t} }.
\end{align}

{\bf JS-GAN:}
The JS divergence without approximation can be formed within the
$f$-GAN framework.
Defining $f(r) = -(r+1) \log \frac{r+1}{2} + r \log r$ gives the JS divergence as follows:
\begin{align}\label{eq:JS-div}
\begin{split}
  \mathcal{D}_{\rm JS} \pt{\Vec{y} \| \Vec{\hat y}} &=
  \frac{1}{2} \int p \pt{\Vec{y}} \log \frac{ 2p\pt{\Vec{y}} }{ p\pt{\Vec{y}} + q\pt{\Vec{\hat y}} } d\Vec{y} \\
  &+
  \frac{1}{2} \int q \pt{\Vec{\hat y}} \log \frac{ 2q\pt{\Vec{\hat y}} }{ p\pt{\Vec{y}} + q\pt{\Vec{\hat y}} } d\Vec{y}.
\end{split}
\end{align}
the discriminator loss
$L_{\rm D}^{(\mathrm{JS}\text{-}\mathrm{GAN})} \pt{\Vec{y}, \Vec{\hat y}}$
is defined as follows:
\begin{align}\label{eq:JSGAN_D}
\begin{split}
L_{\rm D}^{(\mathrm{JS}\text{-}\mathrm{GAN})} \pt{\Vec{y}, \Vec{\hat y}} &=
-\frac{1}{T} \sum_{t=1}^{T} \log \frac{2}{1 + \exp \pt{ - D \pt{\Vec{y}_t} }  }  \\
&\quad -
\frac{1}{T} \sum_{t=1}^{T} \log \pt{ 2 - \frac{2}{1 + \exp \pt{ - D \pt{\Vec{\hat y}_t} } }},
\end{split}
\end{align}
while the adversarial loss
$L_{\rm ADV}^{(\mathrm{JS}\text{-}\mathrm{GAN})} \pt{\Vec{\hat y}}$
is defined as follows:
\begin{align}\label{eq:JSGAN_G}
L_{\rm ADV}^{(\mathrm{JS}\text{-}\mathrm{GAN})} \pt{\Vec{\hat y}} &=
-\frac{1}{T} \sum_{t=1}^{T} \log \frac{2}{1 + \exp \pt{ - D \pt{\Vec{\hat y}_t} }  }.
\end{align}
Note that,
the approximated JS divergence minimized by the original GAN is
$2 \mathcal{D}_{\rm JS}(\Vec{y} \| \Vec{\hat y}) - \log(4) $~\cite{goodfellow14}.

\subsubsection{Wasserstein GAN (W-GAN)~\cite{arjovsky17}}
To stabilize the extremely unstable training of the original GAN,
Arjovsky et al.~\cite{arjovsky17} proposed the W-GAN,
which minimizes the Earth-Mover's distance (Wasserstein-1).
The Earth-Mover's distance is defined as follows:
\begin{align}\label{eq:EMD}
  \mathcal{D}_{\rm EM} \pt{\Vec{y}, \Vec{\hat y}}
  &= \inf_{\gamma} \mathbb{E}_{\pt{\Vec{y}, \Vec{\hat y}} \sim \gamma } \br{ \| \Vec{y} - \Vec{\hat y} \|},
\end{align}
where $\gamma(\Vec{y}, \Vec{\hat y})$
is the joint distribution whose marginals
are respectively the distributions of $\Vec{y}$ and $\Vec{\hat y}$.
On the basis of the Kantorovich--Rubinstein duality~\cite{transport09},
the discriminator loss
$L_{\rm D}^{(\mathrm{W}\text{-}\mathrm{GAN})} \pt{\Vec{y}, \Vec{\hat y}}$
is defined as follows:
\begin{align}\label{eq:WGAN_D}
L_{\rm D}^{(\mathrm{W}\text{-}\mathrm{GAN})} \pt{\Vec{y}, \Vec{\hat y}} &=
-\frac{1}{T} \sum_{t=1}^{T} D \pt{\Vec{y}_t}
+\frac{1}{T} \sum_{t=1}^{T} D \pt{\Vec{\hat y}_t},
\end{align}
while the adversarial loss
$L_{\rm ADV}^{(\mathrm{W}\text{-}\mathrm{GAN})} \pt{\Vec{\hat y}}$
is defined as follows:
\begin{align}\label{eq:WGAN_G}
L_{\rm ADV}^{(\mathrm{W}\text{-}\mathrm{GAN})} \pt{\Vec{\hat y}} &=
-\frac{1}{T} \sum_{t=1}^{T} D \pt{\Vec{\hat y}_t}.
\end{align}
We assume the discriminator to be the
$K$-Lipschitz function.
Namely,
after updating the discriminator,
we clamp its weight parameters to a fixed interval such as $[-0.01, 0.01]$.

\subsubsection{Least Squares GAN (LS-GAN)~\cite{mao17}}
To avoid the gradient vanishing problem of the original GAN using the sigmoid cross entropy,
Mao et al.~\cite{mao17} proposed the LS-GAN,
which formulates the objective function minimizing the mean squared error.
The discriminator loss
$L_{\rm D}^{(\mathrm{LS}\text{-}\mathrm{GAN})} \pt{\Vec{y}, \Vec{\hat y}}$
is defined as follows:
\begin{align}\label{eq:LSGAN_D}
\begin{split}
L_{\rm D}^{(\mathrm{LS}\text{-}\mathrm{GAN})} \pt{\Vec{y}, \Vec{\hat y}} &=
\frac{1}{2T} \sum_{t=1}^{T} \pt{ D \pt{\Vec{y}_t} - b }^2 \\
&\quad+
\frac{1}{2T} \sum_{t=1}^{T} \pt{ D \pt{\Vec{\hat y}_t} - a }^2,
\end{split}
\end{align}
while the adversarial loss
$L_{\rm ADV}^{(\mathrm{LS}\text{-}\mathrm{GAN})} \pt{\Vec{\hat y}}$
is defined as follows:
\begin{align}\label{eq:LSGAN_G}
L_{\rm ADV}^{(\mathrm{LS}\text{-}\mathrm{GAN})} \pt{\Vec{\hat y}} &=
\frac{1}{2T} \sum_{t=1}^{T} \pt{ D \pt{\Vec{\hat y}_t} - c }^2,
\end{align}
where $a$, $b$, and $c$ denote the labels that make
the discriminator recognize
the generated data as generated,
the natural data as natural,
and
the generated data as natural.
When they satisfy the conditions
$b - c = 1$
and
$b - a = 2$,
the divergence to be minimized is the
Pearson $\mathcal{X}^2$
divergence between
$p(\Vec{y}) + q(\Vec{\hat y})$ and $2q(\Vec{\hat y})$.
Because we found that these conditions degrade
quality of synthetic speech,
we used alternative conditions suggested in Eq. (9) of \cite{mao17},
i.e.,
$a = 0$, $b = 1$, and $c = 1$.

\subsection{Discussions}
\label{sec:discussions}
The proposed loss function (Eq.~(\ref{eq:PM})) is the
combination of a multi-task learning algorithm
using discriminators~\cite{huang15}
and GANs.
In defining
$L_{\rm G}(\Vec{y}, \Vec{\hat y}) = L_{\mathrm{ADV}}^{\mathrm{(GAN)}}(\Vec{\hat y})$,
the loss function is equivalent to
that for the GAN.
Comparing with the GANs,
our method is a fully supervised setting,
i.e.,
we utilize the
referred input and output parameters~\cite{reed16gatti}
without a latent variable.
Also,
since only the backpropagation algorithm is used for training,
a variety of DNN architectures such as long short-term memory (LSTM)~\cite{zen15}
can be used as the acoustic models and discriminator.

Using the designed feature function
$\Vec{\phi}(\cdot)$,
we can choose not only analytically derived features (e.g., GV and MS)
but also automatically derived features (e.g., auto-encoded features~\cite{hinton06}).

\begin{figure}[!t]
  \centering
  \includegraphics[width=0.98\linewidth]{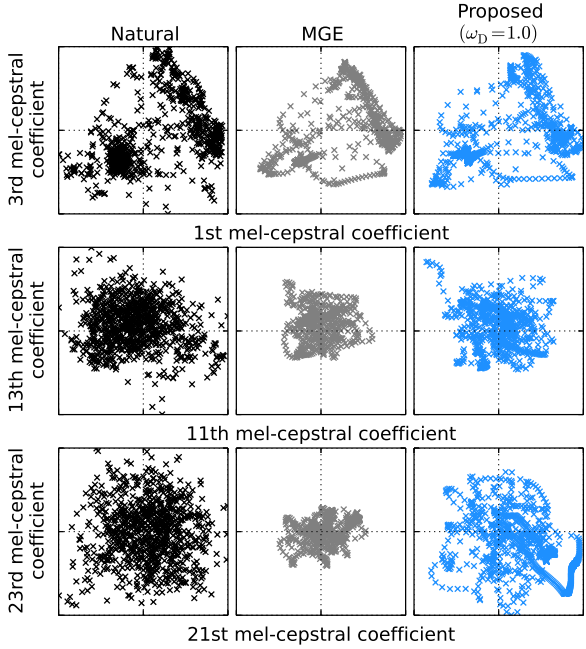}
  \vspace{-20pt}
  \caption{
    Scatter plots of mel-cepstral coefficients with several pairs of dimensions.
    From the left, the figures correspond to natural speech,
    the conventional MGE algorithm,
    and the proposed algorithm $(\omega_{\mathrm{D}} = 1.0)$.
    These mel-cepstral coefficients were extracted from one utterance of the evaluation data.
  }
  \label{fig:sct}
\end{figure}

As described above,
our algorithm
makes the distribution of the generated speech parameters
close to that of the natural speech.
Since we perform generative adversarial training with DNNs,
our algorithm comes to have a more complicated probability distribution
than the conventional Gaussian distribution.
Figure~\ref{fig:sct} plots natural and generated speech parameters
with several mel-cepstral coefficient pairs.
Whereas the parameters of the conventional algorithm are narrowly distributed,
those of the proposed algorithm are
as widely distributed as the natural speech.
Moreover,
we can see that
the proposed algorithm has a greater effect on the distribution of
the higher order of the mel-cepstral coefficients.

\begin{figure}[!t]
  \centering
  \includegraphics[scale=0.45]{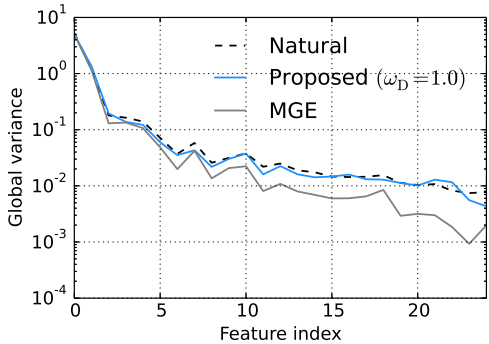}
  \vspace{-10pt}
  \caption{
    Averaged GVs of mel-cepstral coefficients.
    Dashed, black, and blue lines correspond to
    natural speech,
    the conventional MGE,
    and the proposed algorithm, respectively.
  }
  \label{fig:GVs}
\end{figure}

\begin{figure}[!t]
  \centering
  \includegraphics[width=0.98\linewidth]{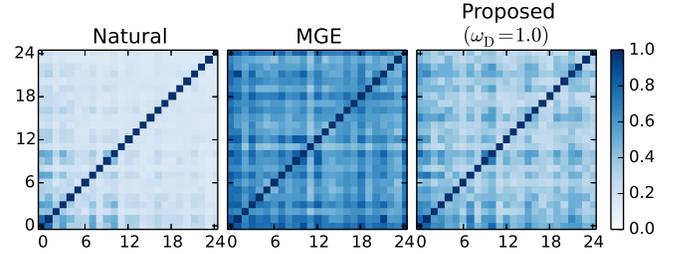}
  \vspace{-60pt}
  \caption{
    MICs of natural and generated mel-cepstral coefficients.
    The MIC ranges from 0.0 to 1.0,
    and the two variables with a strong correlation have a
    value closer to 1.0.
    From the left,
    the figures correspond to natural speech,
    the conventional MGE algorithm,
    and the proposed algorithm $(\omega_{\mathrm{D}} = 1.0)$.
    These MICs were calculated from one utterance of the evaluation data.
  }
  \label{fig:mics}
\end{figure}

Here,
one can explore which components
(e.g., analytically derived features and intuitive reasons~\cite{marco16})
the algorithm changes.
Figure~\ref{fig:GVs} plots the averaged GVs of
natural and generated speech parameters.
We can see that the GV generated by the proposed algorithm
is closer to the natural GV than that of the one produced by
the conventional algorithm.
This is quite natural result because compensating distribution differences
is related to minimizing moments differences~\cite{li15gmmn,goodfellow16}.
Then,
we calculated a maximal information coefficient (MIC)~\cite{rashef11mic}
to quantify a nonlinear correlation among the speech parameters.
The results are shown in Fig.~\ref{fig:mics}.
As reported in \cite{ijima16},
we can see that there are weak correlations among the natural speech parameters,
whereas strong correlations are observed among those of the generated speech parameters
of the MGE training.
Moreover,
the generated mel-cepstral coefficients of our algorithm
have weaker correlations than those of the MGE training.
These results suggest that the proposed algorithm compensates
not only the GV of the generated speech parameters
but also the correlation among the parameters.
Also,
the statistics of continuous $F_0$,
phoneme duration, and mora duration are listed in
Tables \ref{tab:stats_clf0}, \ref{tab:stats_pdur}, and \ref{tab:stats_mdur},
respectively.
The bold values are the closest to natural statistics in the results.
\begin{table}[!t]
\renewcommand{\arraystretch}{1.3}
\caption{
  Statistics of
  natural (``Natural'') and
  generated (``MGE'' and ``Proposed'')
  continuous $F_0$
}
\label{tab:stats_clf0}
\centering
\begin{tabular}{l|rr}
\hline
\hline
 & Mean & Variance \tabularnewline
\hline
Natural & 4.8784 & 0.076853 \tabularnewline
MGE & 4.8388 & 0.032841 \tabularnewline
Proposed ($\omega_{\rm D} = 1.0$) & {\bf 4.8410} & {\bf 0.032968} \tabularnewline
\hline
\hline
\end{tabular}
\end{table}
\begin{table}[!t]
\renewcommand{\arraystretch}{1.3}
\caption{
  Statistics of
  natural (``Natural'') and
  generated (``MSE'' and ``Proposed(*)'')
  phoneme duration
}
\label{tab:stats_pdur}
\centering
\begin{tabular}{l|rr}
\hline
\hline
 & Mean & Variance \tabularnewline
\hline
Natural & 16.314 & 126.20 \tabularnewline
MSE & 14.967 & 47.665 \tabularnewline
Proposed (phoneme, $\omega_{\rm D} = 1.0$) & 14.963 & {\bf 75.471} \tabularnewline
Proposed (mora, $\omega_{\rm D} = 1.0$) & {\bf 15.074} & 73.207 \tabularnewline
\hline
\hline
\end{tabular}
\end{table}
\begin{table}[!t]
\renewcommand{\arraystretch}{1.3}
\caption{
  Statistics of
  natural (``Natural'') and
  generated (``MSE'' and ``Proposed(*)'')
  mora duration
}
\label{tab:stats_mdur}
\centering
\begin{tabular}{l|rr}
\hline
\hline
 & Mean & Variance \tabularnewline
\hline
Natural & 25.141 & 131.93 \tabularnewline
MSE & 23.492 & 60.891 \tabularnewline
Proposed (phoneme, $\omega_{\rm D} = 1.0$) & 24.794 & {\bf 96.828} \tabularnewline
Proposed (mora, $\omega_{\rm D} = 1.0$) & {\bf 24.978} & 96.682 \tabularnewline
\hline
\hline
\end{tabular}
\end{table}
In Tables \ref{tab:stats_pdur} and \ref{tab:stats_mdur},
``Proposed (phoneme)''
and
``Proposed (mora)''
indicate that the proposed methods
applied to phoneme and mora duration, respectively.
We can see that the proposed method also makes the statistics
closer to those of the natural speech than the conventional method.
In the results concerning duration generations,
``Proposed (mora),''
tends to reduce the difference in the mean
rather than in the variance.

\newcolumntype{C}{>{\centering}p{0.25\linewidth}}
\begin{table*}[!t]
\renewcommand{\arraystretch}{1.3}
\caption{
  Architectures of DNNs used in TTS evaluations.
  Feed-Forward networks were used for all architectures.
  ReLU indicates rectified linear unit~\cite{glorot11relu}.}
\label{tab:DNNs}
\centering
\begin{tabular}{c|CCC}
\hline
\hline
 & Spectral parameter generation \\ (sections IV-B-1 and IV-B-2) &  Spectral and $F_0$ parameter generation \\ (section IV-B-3) & Duration generation \\ (section IV-B-4) \tabularnewline
\hline
Acoustic models & 274--3 $\times$ 400 (ReLU)--75 (linear) & 442--3 $\times$ 512 (ReLU)--94 (linear) & 442--3 $\times$ 512 (ReLU)--94 (linear) \tabularnewline
Discriminator & 25--2 $\times$ 200 (ReLU)--1 (sigmoid) & 26--3 $\times$ 256 (ReLU)--1 (sigmoid) & 1--3 $\times$ 256 (ReLU)--1 (sigmoid) \tabularnewline
Duration models & N/A & 439--3 $\times$ 256 (ReLU)--1 (linear) & 439--3 $\times$ 256 (ReLU)--1 (linear) \tabularnewline
\hline
\hline
\end{tabular}
\end{table*}

Our algorithm for spectrum and $F_0$,
proposed in Section III-C,
compensates the joint distribution of them.
Therefore,
we can perform the distribution compensation
considering correlations~\cite{tanaka14_ieice}
between different features.
Also,
compensating dimensionality differences~\cite{kang14}
can be applied for deceiving the discriminator.
Since the time resolutions in phoneme duration and mora duration are different,
our algorithm considering isochrony is related to
multi-resolution GAN~\cite{zhang16}
and hierarchical duration modeling~\cite{yin16}.

Regarding related work,
Kaneko et al.~\cite{kaneko17advps}
proposed a generative adversarial network-based
post-filter for TTS.
The post-filtering process has high portability
because it is independent of original speech synthesis procedures,
but it comes at a high computation cost and
has a heavy disk footprint in synthesis.
In contrast,
our algorithm can directly utilize original synthesis procedures~\cite{zen16fastLSTM}.
Also,
we expect that our algorithm can be extended to waveform synthesis~\cite{tokuda16,oord16wavenet}.

\section{Experimental Evaluation}
In this section,
we evaluate the effectiveness of the proposed algorithm
in terms of
spectral parameters,
$F_0$,
and duration generation
in DNN-based TTS,
and then evaluate spectral parameter conversion in DNN-based VC.

\subsection{Experimental Conditions in TTS Evaluation}
We used speech data of a male speaker taken
from the ATR Japanese speech database~\cite{sagisaka90}.
The speaker uttered 503 phonetically balanced sentences.
We used 450 sentences (subsets A to I) for the training and
53 sentences (subset J) for the evaluation.
Speech signals were sampled at a rate of 16~kHz,
and the shift length was set to 5~ms.
The 0th-through-24th mel-cepstral coefficients were used as spectral parameters
and $F_0$ and 5 band-aperiodicity~\cite{kawahara01, ohtani06}
were used as excitation parameters.
The STRAIGHT analysis-synthesis system~\cite{kawahara99} was
used for the parameter extraction and the waveform synthesis.
To improve training accuracy,
speech parameter trajectory smoothing~\cite{takamichi15blizzard}
with a 50~Hz cutoff modulation frequency was applied to
the spectral parameters in the training data.
In the training phase,
spectral features were normalized to have zero-mean unit-variance,
and 80\% of the silent frames were removed from the training data
in order to increase training accuracy.

The DNN architectures are listed in Table~\ref{tab:DNNs}.
In the spectral parameter generation (sections IV-B-1 and IV-B-2),
the acoustic models predicted static-dynamic feature sequence of
the mel-cepstral coefficients (75-dim.)
from the 274-dimensional linguistic features
frame by frame,
and the discriminator used frame-wise static mel-cepstral coefficients (25-dim.).
Here,
since $F_0$, band-aperiodicity,
and duration of natural speech were directly used
for the speech waveform synthesis,
we only used some of the prosody-related features such as the accent type.
In the spectral parameter and $F_0$ generation (section IV-B-3),
the acoustic models predicted static-dynamic feature sequence of
the mel-cepstral coefficients,
continuous log $F_0$~\cite{yu11},
and band-aperiodicity with a voiced/unvoiced flag (94-dim.)
from the 442-dimensional linguistic features
frame by frame,
and the discriminator used the joint vector of the
frame-wise static mel-cepstral coefficients and continuous log $F_0$ (26-dim.).
In the duration generation (section IV-B-3),
we constructed duration models
that generate phoneme duration from corresponding linguistic features (439-dim).
The acoustic models were trained using MGE training.

In the training phase,
we ran the training algorithm based on minimizing the MSE (Eq.~(\ref{eq:LMSE}))~\cite{zen13dnn}
frame-by-frame for the initialization of acoustic models
and then we ran the conventional MGE training~\cite{wu16mge}
with 25 iterations.
Here,
``iteration'' means using all the training data (450 utterances)
once for training.
The discriminator was initialized using natural
speech and synthetic speech after the MGE training.
The number of iterations for the discriminator initialization was 5.
The proposed training and discriminator re-training were performed with 25 iterations.
The expectation values
$E_{L_{\mathrm{MGE}}}$
and
$E_{L_{\mathrm{ADV}}}$
were estimated at each iteration step.

\subsection{Evaluation in TTS}
\subsubsection{Objective Evaluation with Hyper-parameter Settings}
In order to evaluate our algorithm,
we calculated the parameter generation loss defined in Eq.~(\ref{eq:LG}) and
the spoofing rate of the synthetic speech.
The spoofing rate is the number of spoofing synthetic speech parameters
divided by the total number of synthetic speech parameters
in the evaluation data.
Here,
``spoofing synthetic speech parameter''
indicates a parameter for which the discriminator recognized
the synthetic speech as natural.
The discriminator for calculating the spoofing rates was
constructed using natural speech parameters and generated speech parameters
of the conventional MGE training.
The generation loss and spoofing rates were first calculated with
various hyper-parameter $\omega_{\mathrm{D}}$ settings.

\begin{figure}[tb]
  \centering
  \includegraphics[width=0.98\linewidth]{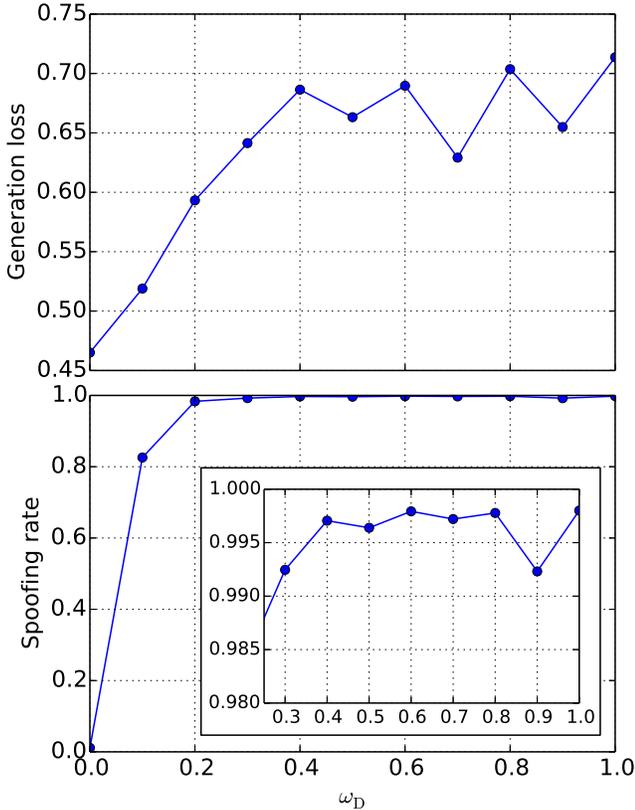}
  \vspace{-20pt}
  \caption{
    Parameter generation loss (above) and spoofing rate (below)
    for various $\omega_{\mathrm{D}}$ for spectral parameter generation in TTS.
  }
  \label{fig:obj}
\end{figure}

Figure~\ref{fig:obj} shows the results for the generation loss and spoofing rate.
As $\omega_{\mathrm{D}}$ increases from 0.0,
the generation loss monotonically increases,
but from 0.4, we cannot see any tendency.
On the other hand,
the spoofing rate significantly increases as
$\omega_{\mathrm{D}}$ increases from 0.0 to 0.2;
from 0.2, the value does not vary much.
These results demonstrate that the proposed training algorithm makes
the generation loss worse but can train the acoustic models to deceive
the discriminator;
in other words,
although our method does not necessarily decrease the generation error,
it tries to reduce the difference between the distributions of
natural and generated speech parameters by taking the adversarial loss
into account during the training.

\subsubsection{Investigation of Convergence in Training}
\begin{figure}[tb]
  \centering
  \includegraphics[width=0.98\linewidth]{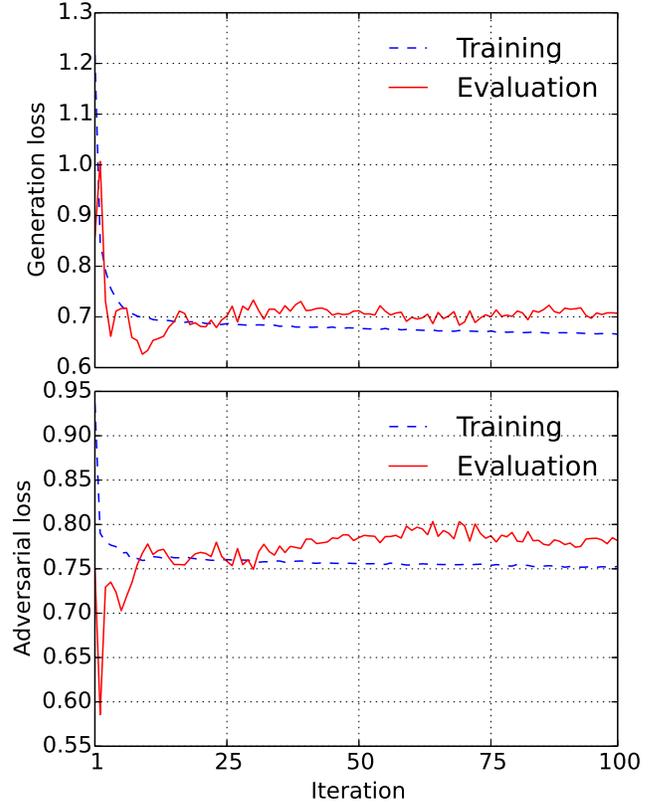}
  \vspace{-10pt}
  \caption{
    Parameter generation loss (above) and adversarial loss (below)
    for the training data (blue-dashed line) and evaluation data(red line).
  }
  \label{fig:conv}
\end{figure}

To investigate the convergence of the proposed training algorithm,
we ran the algorithm through 100 iterations.
Figure~\ref{fig:conv} plots the generation loss and adversarial loss for the
training and evaluation data.
We can see that both loss values are almost monotonically decreased in training.
Although the values of evaluation data strongly vary after a few iterations,
they can converge after several more iterations.

\subsubsection{Subjective Evaluation of Spectral Parameter Generation}
A preference test (AB) test was conducted
to evaluate the quality of speech produced by the algorithm.
We generated speech samples with three methods:
\begin{description}
  \item[MGE:] conventional MGE ($=$ Proposed $(\omega_{\mathrm{D}} = 0.0)$)
  \item[Proposed $(\omega_{\mathrm{D}} = 0.3)$:] \hspace{62pt} spoofing rate $> 0.99$
  \item[Proposed $(\omega_{\mathrm{D}} = 1.0)$:] \hspace{62pt} standard setting
\end{description}
Every pair of synthetic speech samples generated by using
each method was presented to listeners in random order.
Listeners participated in the assessment by using
our crowdsourced subjective evaluation systems.

\begin{figure}[tb]
  \centering
  \includegraphics[width=0.98\linewidth]{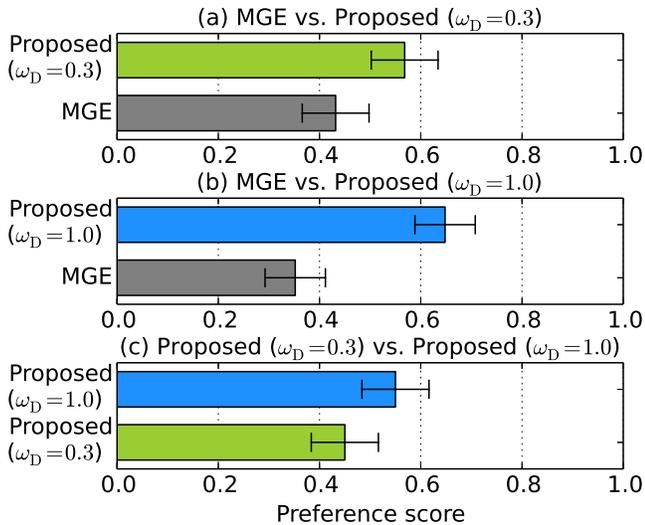}
  \vspace{-10pt}
  \caption{
    Preference scores of speech quality with 95\% confidence intervals
    (spectral parameter generation in TTS).
    From the top,
    the numbers of listeners were
    22, 24, and 22, respectively.
  }
  \label{fig:sbj_tts_sp}
\end{figure}

The results are shown in Fig.~\ref{fig:sbj_tts_sp}.
In Figs.~\ref{fig:sbj_tts_sp}(a) and (b),
the proposed algorithm outperforms conventional
MGE training algorithm in both hyper-parameter settings.
Therefore,
we can conclude that our algorithm
robustly yields significant improvement in terms of speech quality
regardless its hyper-parameter setting.
Henceforth,
we set the hyper-parameter to 1.0 for the
following evaluations
because
Fig.~\ref{fig:sbj_tts_sp} (c) shows that
the score of
``Proposed ($\omega_{\rm D} = 1.0$)''
was slightly better than
that of
``Proposed ($\omega_{\rm D} = 0.3$).''

\subsubsection{Subjective Evaluation of $F_0$ Generation}
We evaluated
the effect of the proposed algorithm for F0 generation. We
conducted a subjective evaluation using the following three
methods:
\begin{description}
  \item[MGE:] conventional MGE
  \item[Proposed (sp):] \hspace{32pt} proposed algorithm applied only to spectral parameters
  \item[Proposed (sp+F0):] \hspace{47pt} proposed algorithm applied to spectral and $F_0$ parameters
\end{description}
Every pair of synthetic speech samples generated by using
each method was presented to listeners in random order.
Since
Fig.~\ref{fig:sbj_tts_sp} has already demonstrated that
the proposed algorithm improves synthetic speech quality
in terms of generating spectral parameters,
we did not compare
``Proposed (sp)''
with
``MGE.''
Listeners participated in the assessment by using
our crowdsourced subjective evaluation systems.

\begin{figure}[tb]
  \centering
  \includegraphics[width=0.98\linewidth]{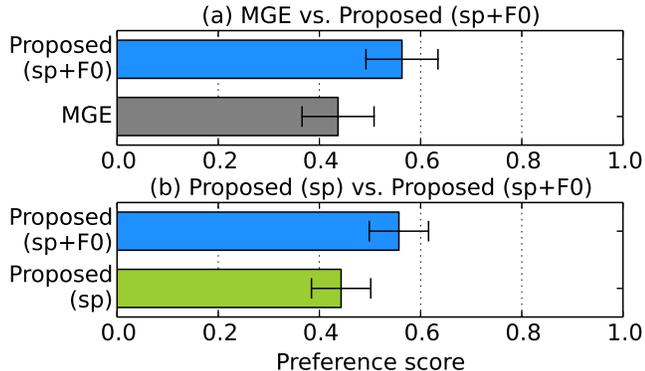}
  \vspace{-10pt}
  \caption{
    Preference scores of speech quality with 95\% confidence intervals
    (spectral parameter and $F_0$ generation in TTS).
    From the top,
    the numbers of the listeners were 19 and 28, respectively.
  }
  \label{fig:sbj_tts_spf0}
\end{figure}

Figure~\ref{fig:sbj_tts_spf0} shows the results.
Since the score of
``Proposed (sp+F0)''
is much higher than those of
``Proposed (sp)''
and
``MGE,''
we can confirm the effectiveness of the proposed
algorithm for not only spectral parameters but also $F_0$.

\subsubsection{Subjective Evaluation of Duration Generation}
We evaluated the effect of the proposed algorithm for duration generation.
We conducted a subjective evaluation using the following three methods:
\begin{description}
  \item[MSE:] conventional MSE
  \item[Proposed (phoneme):] \hspace{54pt} proposed algorithm applied to phoneme duration
  \item[Proposed (mora):] \hspace{47pt} proposed algorithm applied to mora duration
\end{description}
The preference AB test was conducted in the same manner
as in the previous evaluation described in Section.

\begin{figure}[!t]
  \centering
  \includegraphics[width=0.98\linewidth]{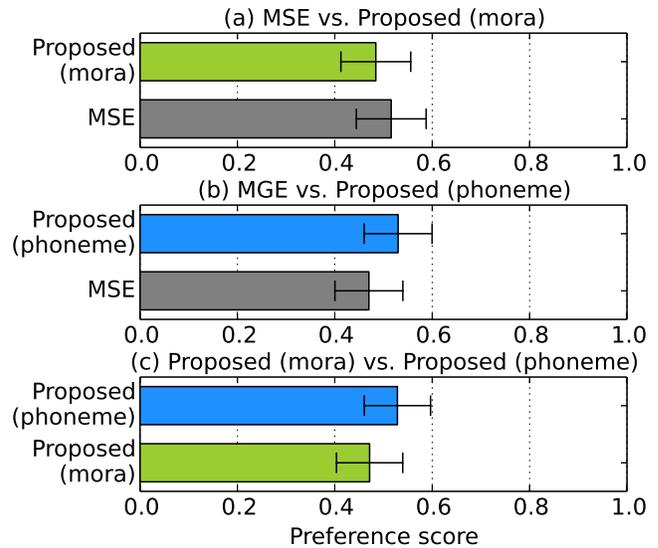}
  \vspace{-10pt}
  \caption{
    Preference scores of speech quality with 95\% confidence intervals
    (duration generation in TTS).
    From the top,
    the numbers of the listeners were 19, 20, and 21, respectively.
  }
  \label{fig:sbj_tts_dur}
\end{figure}

The results are shown in Fig.~\ref{fig:sbj_tts_dur}.
There are no significant differences in
the resulting scores.
To investigate the reason,
we constructed an discriminator that distinguishes
conventional MSE and natural speech,
and calculated the classification accuracy.
We expect that our algorithm works better
when the conventional generated parameters are much distinguished
from the natural ones.
\begin{figure}[!t]
  \centering
  \includegraphics[width=0.98\linewidth]{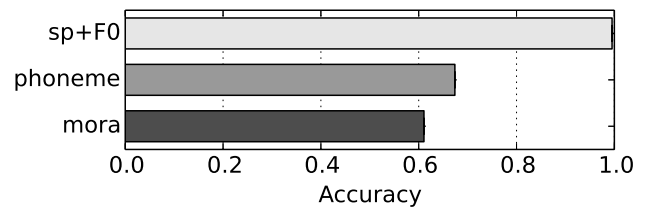}
  \vspace{-10pt}
  \caption{
    Accuracy of discriminator.
    ``sp+F0'', ``phoneme'', and ``mora''
    denote using the spectral parameters and $F_0$,
    phoneme durations,
    and mora durations
    for discriminating the natural and synthetic speech,
    respectively.
  }
  \label{fig:acc_asv}
\end{figure}
As shown in Fig.~\ref{fig:acc_asv},
the accuracy of the discriminator that uses durations is lower than
that of the discriminator that uses spectral parameters and $F_0$.
This result infers that distribution compensation
by our algorithm does not work well in duration generation.
Henceforth,
we did not apply the proposed algorithm for generating durations.

\subsubsection{Comparison to GV Compensation}
Figure~\ref{fig:GVs} demonstrated that
our method compensates the GV of the generated speech parameters.
In addition,
we investigate whether or not our method improves speech quality
more than explicit GV compensation.
We applied the post-filtering process~\cite{toda12}
to the spectral and $F_0$ parameters generated by the MGE training.
A preference AB test with 29 listeners was conducted by using
our crowd-sourced subjective evaluation systems.

\begin{figure}[!t]
  \centering
  \includegraphics[width=0.98\linewidth]{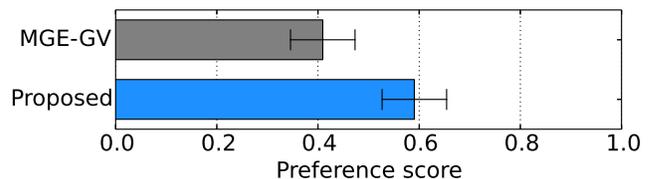}
  \vspace{-10pt}
  \caption{
    Preference scores of speech quality with 95\% confidence intervals
    (compared with the GV compensation).
  }
  \label{fig:sbj_tts_gvpf}
\end{figure}

Figure~\ref{fig:sbj_tts_gvpf} shows the results.
Since the score of ``Proposed'' is higher than
that of the conventional GV post-filter (``MGE-GV''),
we can conclude that
our method produces more gain in speech quality than the conventional GV compensation.

\subsubsection{Effect of Feature Function}
We investigate whether the feature function used in
anti-spoofing is effective to our method.
We adopted the following two functions:
\begin{description}
  \item[Identity:] \hspace{6pt} $\Vec{\phi}(\Vec{y}) = \Vec{y}$
  \item[Static \& delta~\cite{sahidullah15}:] \hspace{52pt} $\Vec{\phi}(\Vec{y}) = \Vec{W}\Vec{y}$
\end{description}
``Identity'' is equivalent to not using the feature function.
When ``Static \& delta'' is adopted,
joint vectors of the static, delta,
and delta-delta mel-cepstral coefficients
and continuous $F_0$ are input to the discriminator.
A preference AB test with 31 listeners was conducted by using
our crowd-sourced subjective evaluation systems.

\begin{figure}[!t]
  \centering
  \includegraphics[width=0.98\linewidth]{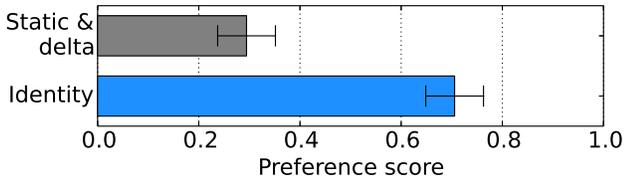}
  \vspace{-10pt}
  \caption{
    Preference scores of speech quality with 95\% confidence intervals
    (effect of the feature function which is used in anti-spoofing).
  }
  \label{fig:sbj_tts_delta}
\end{figure}

Figure~\ref{fig:sbj_tts_delta} shows the results.
Clearly, the score of ``Static \& delta''
is much lower than that of ``Identity.''
From this result,
although ``Static \& delta''
effectively distinguishes natural and synthetic speech,
it does not improve speech quality.

\subsubsection{Subjective Evaluation Using Complicated Model Architecture}
Only simple Feed-Forward networks were used in the above-described evaluations.
Accordingly,
we confirm whether our method can improve speech quality
even when more complicated networks are used.
We used two-layer uni-directional LSTMs~\cite{zen15}
as both acoustic models and discriminator.
The numbers of memory cells in the acoustic models and discriminator were 256 and 128,
respectively.
Our method was applied to spectral and $F_0$ parameters.
MGE (``MGE'') and the proposed (``Proposed'') training algorithm were compared.
A preference AB test with 19 listeners was conducted by using
our crowd-sourced subjective evaluation systems.

\begin{figure}[!t]
  \centering
  \includegraphics[width=0.98\linewidth]{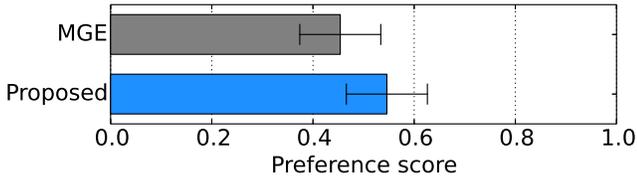}
  \vspace{-10pt}
  \caption{
    Preference scores of speech quality with 95\% confidence intervals
    (comparison in using LSTM).
  }
  \label{fig:sbj_tts_lstm}
\end{figure}

Figure~\ref{fig:sbj_tts_lstm} shows the results.
Since the score of ``Proposed'' is higher than
that of ``MGE,''
we can demonstrate that our method works for
not only simple architectures but also complicated ones.

\subsubsection{Effect of Divergence of GAN}
As the final investigation regarding TTS,
we compared speech qualities of various GANs.
We adopted the following GANs:
\begin{description}
  \item[GAN:] Eqs. (\ref{eq:GAN_D}) and (\ref{eq:GAN_G})
  \item[KL-GAN:] \hspace{12pt} Eqs. (\ref{eq:KLGAN_D}) and (\ref{eq:KLGAN_G})
  \item[RKL-GAN:] \hspace{18pt} Eqs. (\ref{eq:RKLGAN_D}) and (\ref{eq:RKLGAN_G})
  \item[JS-GAN:] \hspace{10pt} Eqs. (\ref{eq:JSGAN_D}) and (\ref{eq:JSGAN_G})
  \item[W-GAN:] \hspace{8pt} Eqs. (\ref{eq:WGAN_D}) and (\ref{eq:WGAN_G})
  \item[LS-GAN:] \hspace{10pt} Eqs. (\ref{eq:LSGAN_D}) and (\ref{eq:LSGAN_G})
\end{description}
We conducted a MOS test on speech quality.
The synthetic speech generated by using each GAN
was presented to listeners in random order.
55 listeners participated in the assessment by using
our crowdsourced subjective evaluation systems.

\begin{figure}[!t]
  \centering
  \includegraphics[width=0.98\linewidth]{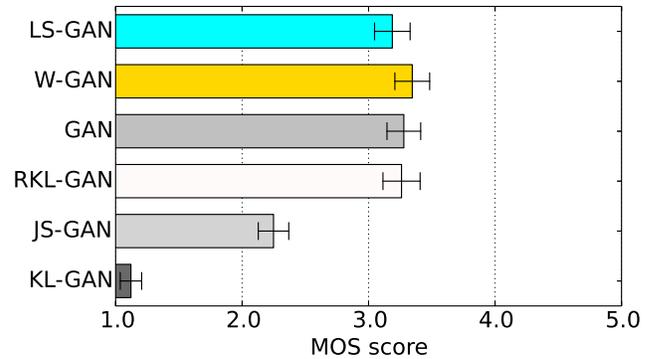}
  \vspace{-10pt}
  \caption{
    MOS scores of speech quality with 95\% confidence intervals
    (comparison in divergences of GANs).
  }
  \label{fig:sbj_tts_gans}
\end{figure}

Figure~\ref{fig:sbj_tts_gans} shows the results.
We can see that our method works
in the case of all divergences except ``KL-GAN'' and ``JS-GAN.''
Two points are noteworthy:
1) minimizing KL-divergence (KL-GAN) did not improve synthetic speech quality,
but the reversed version (RKL-GAN) worked,
and 2) JS-divergence did not work well,
but the approximated version (GAN) worked.
The best GAN in terms of synthetic speech quality was the W-GAN,
whose MOS score was significantly higher than those of the LS-GAN, JS-GAN, and KL-GAN.

\subsection{Experimental Conditions in VC Evaluation}
The experimental conditions such as
dataset used in the evaluation,
speech parameters,
pre-processing of data,
and training procedure were the same as the previous evaluations
except for the dimensionality of spectral parameters and DNN architectures.
We constructed DNNs for male-to-male conversion
and male-to-female conversion.
The hidden layers of the acoustic models and discriminator had
3 $\times$ 512 units
and
3 $\times$ 256 units,
respectively.
The 1st-through-59th mel-cepstral coefficients were converted.
The input 0th mel-cepstral coefficients were directly used
as those of the converted speech.
$F_0$ was linearly transformed,
and band-aperiodicity was not transformed.
Dynamic time warping was used to align total frame lengths of the
input and output speech parameters.

We generated speech samples with
the conventional MGE training
and
the proposed training algorithm.
We conducted a preference AB test to evaluate the converted speech quality.
We presented every pair of converted speech of the two sets in random order
and had listeners select the speech sample that sounded better in quality.
Similarly,
an XAB test on the speaker individuality was conducted
using the natural speech as a reference ``X.''
Eight listeners participated in
assessment of male-to-male conversion case,
and 27 listeners participated in
assessment of male-to-female conversion case using
our crowdsourced subjective evaluation systems.

\subsection{Subjective evaluation in VC}
\begin{figure}[tb]
  \centering
  \includegraphics[width=0.98\linewidth]{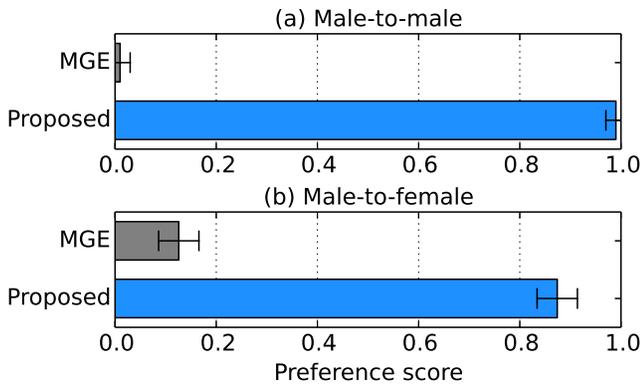}
  \vspace{-10pt}
  \caption{
      Preference scores of speech quality with 95\% confidence intervals
      (DNN-based VC).
  }
  \label{fig:sbj_vc_qual}
\end{figure}

\begin{figure}[tb]
  \centering
  \includegraphics[width=0.98\linewidth]{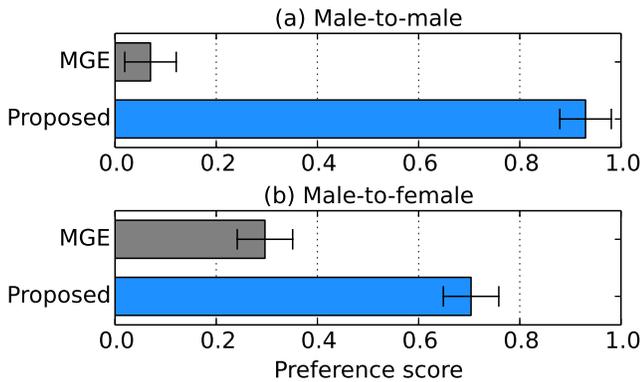}
  \vspace{-10pt}
  \caption{
      Preference scores of speaker individuality with 95\% confidence intervals
      (DNN-based VC).
  }
  \label{fig:sbj_vc_indiv}
\end{figure}

The results of the preference tests on speech quality
and speaker individuality are shown in
Fig.~\ref{fig:sbj_vc_qual} and Fig.~\ref{fig:sbj_vc_indiv},
respectively.
We can find that our algorithm achieves better scores
in speech quality the same as the TTS evaluations.
Moreover,
we can see that the proposed algorithm also improves speaker individuality.
We expect that the improvements are caused by compensating GVs of the
generated speech parameters
which affect speaker individuality~\cite{toda07_MLVC}.
These improvements were observed not only in the inter-gender case
but also cross-gender case.
Therefore,
we have also demonstrated the effectiveness of the algorithm in DNN-based VC.

\section{Conclusion}
In this paper,
we proposed a novel training algorithm for
deep neural network (DNN)-based
high-quality statistical parametric speech synthesis.
The algorithm incorporates a framework of
generative adversarial networks (GANs),
which adversarily train generator networks and discriminator
networks.
In the case of proposed algorithm,
acoustic models of speech synthesis are trained to
deceive the discriminator that distinguishes natural and synthetic speech.
Since the GAN framework minimizes the difference in distributions of
natural and generated data,
the acoustic models are trained to not only minimize
the generation loss but also make the parameter
distribution of the generated speech parameters
close to that of natural speech.
This is a pioneering method of GAN-based speech synthesis and
can be applied not only statistical parametric approaches
but also the ones such as glottal waveform synthesis~\cite{bollepalli17}.
We found that our algorithm compensated
not only global variance
but also correlation among generated speech parameters.
Experimental evaluations were conducted in both
DNN-based text-to-speech (TTS) synthesis and
voice conversion (VC).
The results demonstrate that
the proposed algorithm yields significant improvements
in terms of speech quality in both TTS and VC
regardless of its hyper-parameter settings.
We also found that the proposed algorithm incorporating
the Wasserstein GAN improved synthetic speech quality the most
in comparison with various GANs.
In future work,
we will further investigate the behavior in relation to the hyper-parameter settings,
adopt feature functions which are more effective to detect synthetic speech
than the identity function,
and devise discriminator models with
linguistic~\cite{reed16gatti} dependencies.


%

\ifCLASSOPTIONcaptionsoff
  \newpage
\fi



%
\bibliographystyle{IEEEbib}
\bibliography{tts}

%

\begin{IEEEbiography}{Yuki Saito}
received a B.E. in engineering from
National Institution for Academic Degrees and
Quality Enhancement of Higher Education in 2016.
He is currently working toward an M.E.
in Information Science and Technology
at Graduate School of Information Science and Technology,
the University of Tokyo, Tokyo.
His research interests include
statistical parametric speech synthesis,
machine learning, and machine intelligence.
He received the 14th Best Student Presentation Award of ASJ
and the 2017 IEICE ISS Student Poster Award.
He is a Student Member of the Acoustical Society of Japan.
\end{IEEEbiography}

\begin{IEEEbiography}{Shinnosuke Takamichi}
received his B.E. from Nagaoka University of Technology, Japan, in 2011
and his M.E. and Ph.D. from the Graduate School of Information Science,
Nara Institute of Science and Technology (NAIST),
Japan, in 2013 and 2016, respectively.
He was a short-term researcher at NICT, Kyoto, Japan in 2013,
a visiting researcher of Carnegie Mellon University (CMU) in the-United States
from 2014 to 2015, and a Research Fellow (DC2) of
Japan Society for the Promotion of Science,
from 2014 to 2016.
He is currently a Project Research Associate of the University of Tokyo.
He has received more than ten paper/achievement awards including
the 8th Outstanding Student Paper Award from IEEE Japan Chapter SPS
and the Itakura Prize Innovative Young Researcher Award.
He is a member of ASJ, IPSJ, ISCA and IEEE SPS.
\end{IEEEbiography}


\begin{IEEEbiography}{Hiroshi Saruwatari}
  received B.E., M.E., and Ph.D. degrees
  from Nagoya University, Nagoya, Japan
  in 1991, 1993, and 2000, respectively.
  He joined the Intelligent System Laboratory,
  SECOM Co., Ltd., Tokyo, Japan in 1993,
  where he was engaged in research on the
  ultrasonic array system for acoustic imaging.
  He is currently a Professor at Graduate School of
  Information Science and Technology, the University of Tokyo, Tokyo.
  His research interests include noise reduction,
  array signal processing, blind source separation,
  and sound field reproduction.
  He received paper awards from the IEICE in 2001 and 2006,
  from the Telecommunications Advancement Foundation in 2004 and 2009,
  and at the IEEE-IROS2005 in 2006.
  He won the first prize at the IEEE MLSP2007 Data Analysis Competition for BSS.
  He is a Member of the IEICE, Japan VR Society, and the ASJ.
\end{IEEEbiography}




\end{document}